\documentclass[preprint,letterpaper,amsmath,amssymb]{revtex4}

\usepackage{graphicx}
\usepackage{hyperref}

\newcommand{\ve}{\varepsilon}
\newcommand{\vp}{\varphi}
\newcommand{\pv}[1]{{-  \hspace {-4.0mm} #1}}

\begin{document}


\title{Solutions of coupled BPS equations for two-family Calogero and matrix models}

\author{Velimir Bardek}
\email{bardek@irb.hr}
\author{Stjepan Meljanac}
\email{meljanac@irb.hr}
\author{Daniel Meljanac}
\email{dmeljan@irb.hr}
\affiliation{Institute Rudjer Bo\v{s}kovi\'{c}, Bijeni\v{c}ka cesta 54, HR-10002 Zagreb, Croatia}

\begin{abstract}

We consider a large N, two-family Calogero and matrix model in the Hamiltonian, collective-field approach. The Bogomol'nyi limit appears and the solutions to the coupled Bogomol'nyi-Prasad-Sommerfeld equations are given by the static soliton configurations. We find all solutions close to constant and construct exact one-parameter solutions in the strong-weak dual case. Full classification of these solutions is presented. 

\end{abstract}


\maketitle


\section{\label{sec:1} Introduction}

Recently, a specific duality-based generalization of the Hermitian matrix model has been analyzed \cite{AJJ:2005}. The existence of two collective fields allows one to interpret the generalized matrix model as a two-family Calogero model, implying their equivalence in the collective-field approach. The soliton sector of these models was studied in Refs. \cite{AJJ:2005,AJJ:2006}. The multivortex solutions of the coupled Bogomol'nyi-Prasad-Sommerfeld (BPS) equations were interpreted as giant gravitons \cite{Bena:2005}. Moreover, the multivortex singular solutions have motivated the authors of Refs. \cite{AJJ:2005} and\cite{AJJ:2006} to propose a realization of open-closed string duality \cite{Berenstein}.  The same BPS equations were studied in Refs. \cite{BMEPL:2005}, \cite{BMJHEP:2005}, \cite{BMPRD:2007} and no multivortex solutions were found. 

The purpose of this paper is to give a systematic perturbative method for solving coupled BPS equations and find all solutions close to the constant solution. In addition,  we present the construction of all one-parameter exact solutions which can be continuously connected  to constant solution, in the strong-weak dual variant of the model.

The outline of the paper is as follows. In Sec. \ref{sec:2}, we briefly sketch
the collective-field derivation of coupled BPS equations of the two-family Calogero
model. In Sec. \ref{sec:3}, we propose a perturbative method to find the periodic solutions close to constant. In
Sec. \ref{sec:4}, we present the construction of a one-parameter solutions. Sec. \ref{sec:5}, is devoted to detailed classification of all solutions, which can be continuously connected with the constant solution, in terms of a new, more suitable parameter. Finally, in Sec. \ref{sec:6}, we present our discussion and conclusions.


\section{\label{sec:2} Two-family models in the collective-field approach}
The Hamiltonian of the two-family Calogero system \cite{BFM:2007} reads 
\begin{equation}\label{H}
\begin{split}
H& =-\frac{1}{2m_1}\sum_{i=1}^{N_1}\frac{\partial^2}{\partial x_i^2}+\frac{\lambda_1(\lambda_1-1)}{2m_1}\sum_{i \ne j}^{N_1}\frac{1}{(x_i-x_j)^2} \\
& -\frac{1}{2m_2}\sum_{\alpha=1}^{N_2}\frac{\partial^2}{\partial x_\alpha^2}+
\frac{\lambda_2(\lambda_2-1)}{2m_2}\sum_{\alpha \ne \beta}^{N_2}\frac{1}{(x_\alpha-x_\beta)^2} \\
& +\frac{1}{2}\bigg(\frac{1}{m_1}+\frac{1}{m_2}\bigg)\lambda_{12}(\lambda_{12}-1)
\sum_{i=1}^{N_1}\sum_{\alpha=1}^{N_2}\frac{1}{(x_i-x_\alpha)^2}.
\end{split}
\end{equation}
Here, the first family contains $N_1$ particles of mass $m_1$ at positions $x_i,\ i=1,2, \ldots, N_1$, and the second one contains $N_2$ particles of mass $m_2$ at positions $x_\alpha,\ \alpha=1,2, \ldots, N_2$. All particles interact via two-body inverse-square potentials.

The interaction strength between particles of the first and the second families is parametrized by $\lambda_{12}$. The interaction strengths within each family are parametrized by the coupling constants $\lambda_1$ and $\lambda_2$, respectively. In Eq. \eqref{H}, we imposed the restriction that there be no three-body interactions, which requires \cite{MMS:2003},\cite{Forrester}, \cite{Sen} 
\begin{equation}\label{lambda}
\frac{\lambda_1}{m_1^2}=\frac{\lambda_2}{m_2^2}=\frac{\lambda_{12}}{m_1m_2}.
\end{equation}

It follows from \eqref{lambda} that
\begin{equation}
\lambda_{12}^2=\lambda_1 \lambda_2.
\end{equation}
In addition, we restrict ourselves to the so-called strong-weak dual variant of the model in which the coupling parameters are related by 
\begin{equation}\label{coupling}
\lambda_1 \lambda_2=\lambda_{12}=1.
\end{equation}

In Ref. \cite{BMEPL:2005} we studied the collective-field theory of the two-family Calogero model given by Hamiltonian \eqref{H}. The corresponding collective Hamiltonian is
\begin{equation}\label{H2}
\begin{split}
H_{coll}& = \frac{1}{2m_1}\int dx \rho_1(x)\Big(\partial_x \pi_1(x)\Big)^2 \\
& +\frac{1}{2m_1}\int dx \rho_1(x)\bigg(\frac{\lambda_1-1}{2}\frac{\partial_x\rho_1}{\rho_1}-
\lambda_1 \pi \rho_1^H(x)-\lambda_{12} \pi \rho_2^H(x)\bigg)^2 \\
& +\frac{1}{2m_2}\int dx \rho_2(x)\Big(\partial_x \pi_2(x)\Big)^2 \\
& +\frac{1}{2m_2}\int dx \rho_2(x)\bigg(\frac{\lambda_2-1}{2}\frac{\partial_x\rho_2}{\rho_2}-
\lambda_2 \pi \rho_2^H(x)-\lambda_{12} \pi \rho_1^H(x)\bigg)^2,
\end{split}
\end{equation}
which is a straightforward generalization of \eqref{H}. Here $\rho_1$ and $\rho_2$ are the collective density fields of the first and the second family, respectively, and $\pi_1$ and $\pi_2$ are their conjugate momenta. The collective fields $\rho_1$ and $\rho_2$ are normalized as 
\begin{equation}\label{normalize}
\int dx \rho_1(x)=N_1, \quad \int dx \rho_2(x)=N_2.
\end{equation}
Designation $\rho^H(x)$ denotes the Hilbert transform of $\rho(x)$:
\begin{equation}
    \rho^H(x) = \frac{1}{\pi} \pv \int \frac{dy \rho(y)}{y-x}.
\end{equation}

The Hamiltonian \eqref{H2} is essentially the sum of two positive terms. Its zero-energy classical solutions are zero-momentum, and therefore time-independent, configurations of the collective fields, which are also solutions of the coupled BPS equations
\begin{align}
\frac{\lambda_1-1}{2}\frac{\partial_x\rho_1}{\rho_1}-
\lambda_1 \pi \rho_1^H(x)-\lambda_{12} \pi \rho_2^H(x) & = 0 \label{BPS1}\\
\frac{\lambda_2-1}{2}\frac{\partial_x\rho_2}{\rho_2}-
\lambda_2 \pi \rho_2^H(x)-\lambda_{12} \pi \rho_1^H(x) & = 0. \label{BPS2}
\end{align}

Finding the general exact solutions of these coupled equations for arbitrary couplings and masses is still an open problem, which we briefly discuss  in Sec. \ref{sec:6}. Our main interest here is simpler, and concerns investigating whether the above pair of equations is exactly solvable for a special choice of parameters \eqref{coupling}.

We observe that $\rho_1$ and $\rho_2$ are connected in a very simple way. Namely, by multiplying the second equation \eqref{BPS2} with $\lambda_1$ and subtracting it from the first equation \eqref{BPS1}, we easily get
\begin{equation}
\partial_x \ln (\rho_1\rho_2)=0, \quad \textrm{or} \quad \rho_1\rho_2=c,
\end{equation}
where $c$ is some positive constant. This is a direct consequence of the special choice \eqref{coupling}.

Now, eliminating $\rho_2$ in terms of $\rho_1$ we can concentrate only on the first equation \eqref{BPS1}:
\begin{equation}\label{BPS1a}
\frac{\lambda_1-1}{2}\frac{\partial_x\rho_1}{\rho_1}=
\lambda_1 \pi \rho_1^H(x)+ \pi c \bigg(\frac{1}{\rho_1}\bigg)^H,
\end{equation}
which upon introducing abbreviations $a = \dfrac{2\lambda_1 \pi}{\lambda_1-1}$ and $b = \dfrac{2c\pi}{\lambda_1-1}$ takes the form
\begin{equation}\label{lnrho}
    \frac{d}{dx} \ln \rho(x)=a \rho^H(x) +
    b\Big(\frac{1}{\rho(x)}\Big)^H.
\end{equation}


\section{\label{sec:3}Solutions close to constant}

By having in mind the fact that the Hilbert transform of a constant is zero \cite{Bracwell:99}, it is evident that there always exists a uniform solution of Eq. \eqref{lnrho}:
\begin{equation}\label{const}
\rho(x)=\rho_0.
\end{equation}

Now we are going to construct the most general solutions of Eq. \eqref{lnrho} which are continuously connected with constant solution \eqref{const}.

They are of the form
\begin{equation}\label{expansion}
    \rho(x)=\rho_0 \sum_{i=0}^{\infty} \ve^i \vp_i(x),
\end{equation}
where $\ve$ is a small parameter ($\ve \ll 1$) and $\vp_0(x)=1$.

Expanding $\displaystyle{\frac{d}{dx} \ln \rho}$ and $\displaystyle{\left(\frac{1}{\rho}\right)^H}$ in powers of small parameter $\ve$ we find the infinite set of recursive equations for $\vp_i(x)$ of the form:
\begin{subequations}\label{recursive}
\begin{gather}
   \vp_1'=k_0 \vp_1^H \\
   \vp_2'=k_0 \vp_2^H +
   \frac{1}{2}(\vp_1^2)'+\frac{b}{\rho_0}(\vp_1^2)^H \\
   \vp_3'=k_0 \vp_3^H + (\vp_1\vp_2)'+\frac{2b}{\rho_0}(\vp_1\vp_2)^H-
   \frac{1}{3}(\vp_1^3)'-\frac{b}{\rho_0}(\vp_1^3)^H \\ 
   \vdots \nonumber
\end{gather}
\end{subequations}
where 
\begin{equation}
k_0=a \rho_0 - \frac{b}{\rho_0}
\end{equation}
and the prime denotes derivative with respect to $x$. The solutions up to the second order in $\ve$ expansion are given (up to a phase) by
\begin{subequations}\label{e16}
\begin{align}
\vp_1(x) &= c_1 \cos k_0x \label{e16a}\\
\vp_2(x) &= c_2 \cos 2k_0x + \tilde c_1 \cos k_0x \label{e16b}
\end{align}
\end{subequations}
where constants $c_1$ and $c_2$ are interrelated via
\begin{equation}\label{e17}
c_2=\frac{a \rho_0}{2k_0}c_1^2,
\end{equation}
and $c_1$, and $\tilde c_1$ are arbitrary. Note that $\tilde c_1$ can be removed by redefinition of parameter $\ve \to \tilde \ve = \ve + \dfrac{\tilde c_1}{c_1} \ve^2$, up to the second order in $\ve$.
 
In obtaining $\vp_1$ and $\vp_2$ we have used well-known Hilbert transform formulas \cite{Bracwell:99}
\begin{subequations}
\begin{align}
(\cos kx)^H & =-\mathrm{sign}\, k \sin kx  \\
(\sin kx)^H & =\mathrm{sign}\, k \cos kx .
\end{align}
\end{subequations}

Relations \eqref{e16a} and \eqref{e16b} suggest that $\rho(x)$ is a periodic function. Therefore, we restrict our search for solutions $\rho(x)$ to periodic functions. In this respect, we accommodate our expansion \eqref{expansion}: 
\begin{equation}\label{expansion2}
    \rho(x)=\rho_0 \sum_{i=0}^{\infty} \ve^i C_i \cos ikx.
\end{equation}
Inserting this expansion into the infinite set of recursive relations \eqref{recursive}, we get
\begin{subequations}
\begin{align}
C_i & = \sum_{m=0}^{\infty} C_{i,2m} \ve^{2m} \,  \\
k & = \sum_{m=0}^{\infty} k_{2m} \ve^{2m} \, \label{kexp} \\
C_{i,0} & = C_{1,0}\left(\frac{C_{1,0}}{2k_0}a\rho_0\right)^{i-1}. 
\end{align}
\end{subequations}
All other periodic solutions can be brought to the form \eqref{expansion2} by a suitable change of parameter $\ve$.

Let us now define truncated function
\begin{equation}
    \rho_n(x)=\rho_0 \sum_{i=0}^{n} \ve^i [C_i \cos ikx]_{(n-i)},
\end{equation}
where an index in parentheses denotes the order of the polynomial in parameter $\ve$. Then, it is easy to see that
\begin{equation}
    \ve^n\vp_n(x)=\rho_n(x) - \rho_{n-1}(x). 
\end{equation}

In this way we can rederive relations \eqref{e16a} and \eqref{e16b} and obtain, for example, $\vp_3(x)$:
\begin{align}
    \ve^3\vp_3(x) & =\ve^3 C_{3,0} \cos 3k_0x + \ve C_{1,0} [\cos (k_0+k_2\ve^2)x]_{(2)} \nonumber \\
&+ \ve [C_{1,2}\ve^2-C_{1,0}] \cos k_0x, 
\end{align}
where
\begin{align}
C_{3,0} & = C_{1,0}\left(\frac{C_{1,0}}{2k_0}a\rho_0\right)^{2}  \\
k_2 & = \frac{1}{2}k_0\left(\frac{b}{a\rho_0^2}\right)^2\left(\frac{C_{1,0}}{2k_0}a\rho_0\right)^{2} 
\end{align}

In the next section, we shall give the analytic form for the periodic solution $\rho(x),\ k>0$, and the aperiodic solutions in the limit $k \to 0_+$. 

Finally, we note that in the special case $b=0$, the expansion \eqref{expansion2} can be easily summed. Namely, in this case
\begin{equation}
    k=k_0=a\rho_0,
\end{equation}
and choosing $C_{1,0}=1$ we find $$C_i=2^{1-i},\ i\ge 1,$$
\begin{align}
    \rho(x) &= \rho_0 [1+ \sum_{i=1}^{\infty} \frac{\ve^i}
{2^{i-1}} c_i \cos ikx] ,  \\
		&= \rho_0\frac{\sqrt{1-e^2}}{1-e\cos k_0x},\quad e=\frac{\ve}{1+\frac{\ve^2}{4}} \le 1. \label{e29}
\end{align}
Note that the expansion in $e$ is different from the expansion in $\ve$, but they are both of the type given by Eq. \eqref{expansion2} and describe the same solution. The one-parameter solutions given in \eqref{e29} coincide with corresponding solutions in Refs. \cite{AJJ:2006}, \cite{BMPRD:2007}, and \cite{BFM:2008}.


\section{\label{sec:4}Construction of one-parameter solutions}

Let us now show that there exists a particular one-parameter family of solutions to \eqref{lnrho} which is not necessarily close to constant. To this end we define a family of functions $R_\ve(x)$ depending on parameter $\ve$ and wave vector $k$:
\begin{equation}
    R_\ve(x)=\frac{\sqrt{1-\ve^2}}{1-\ve \cos kx}, 
\end{equation}
where $|\ve| \le 1,\ \ve \in \mathbf{R}$, and $k > 0$.

These functions satisfy
\begin{equation}
    \overline{R_\ve(x)}=\lim_{L \to \infty} \frac{1}{2L} \int_{-L}^{L} dx R_\ve(x)=1.
\end{equation}
We can easily verify that $R_\ve$ can be written as
\begin{equation}\label{Reps}
    R_\ve(x)=1+\frac{\gamma e^{ikx}}{1-\gamma e^{ikx}}+
\frac{\gamma e^{-ikx}}{1-\gamma e^{-ikx}}=
\frac{1-\gamma^2}{1+\gamma^2-2\gamma\cos kx},
\end{equation}
where $|\gamma|<1,\ \gamma\in\mathbf{R}$, and $\displaystyle{\ve=\frac{2\gamma}{1+\gamma^2}}$.

Since function $\displaystyle{\frac{e^{ikx}}{1-\gamma e^{ikx}}}$ is analytic in the $x$-upper-half-plane for $k>0$, and vanishes at infinity in that half-plane, then
\begin{equation}\label{Reps1}
    \left(\frac{e^{ikx}}{1-\gamma e^{ikx}}\right)^H=i
\frac{e^{ikx}}{1-\gamma e^{ikx}}.
\end{equation}
Similarly, we have 
\begin{equation}\label{Reps2}
    \left(\frac{e^{-ikx}}{1-\gamma e^{-ikx}}\right)^H=-i
\frac{e^{-ikx}}{1-\gamma e^{-ikx}}.
\end{equation}
Making use of \eqref{Reps},\eqref{Reps1}, and \eqref{Reps2} we obtain 
\begin{equation}
    (R_\ve)^H=-\frac{\ve \sin kx}{1- \ve \cos
    kx}= \frac{1}{k}(\ln R_\ve)'.
\end{equation}
This is nothing but the one-family Calogero model BPS equation. Namely, for $b=0$, Eq. \eqref{lnrho} reduces to
\begin{equation}
\frac{\rho'}{\rho}=a \rho^H
\end{equation}
implying that corresponding one-parameter solutions are in fact given by 
\begin{equation}
\rho(x)=\rho_0 R_\ve(x),
\end{equation}
where $k=a\rho_0 > 0$.

We are now in a position to construct a class of one-parameter solutions of coupled BPS equations \eqref{BPS1} and \eqref{BPS2}. Let us proceed in a few steps. First note that
\begin{equation}
\left(\frac{R_\ve}{R_\eta}\right)^H = -
    \frac{R_\ve}{R_\eta}\left(\frac{R_\eta}{R_\ve}\right)^H,
\end{equation}
\begin{equation}\label{Reta}
R_\eta(x)=\frac{\sqrt{1-\eta^2}}{1-\eta\cos kx}\,, 
\end{equation}
where $|\eta| \le 1,\ \eta \in \mathbf{R}$.
Using this important relation we construct a general solution of Eq. \eqref{lnrho} given by two equivalent forms
\begin{equation}\label{general}
    \rho(x)=s \frac{R_\ve(x)}{R_\eta(x)}=r_0
    R_\ve(x)+ \alpha \,, 
\end{equation}
where $s>0,\ r_0, \alpha \in \mathbf{R},$ and $\overline{\rho(x)} = \rho_0$.

Namely, after inserting solution \eqref{general} into Eq. \eqref{lnrho}, we find
\begin{equation}
    (\ln \rho)'=k[R_\ve-R_\eta]^H=
    \frac{as}{\sqrt{1-\eta^2}}\left(1-\frac{\eta}{\ve}\right)R_\ve^H+
    \frac{b}{s \sqrt{1-\ve^2}}\left(1-\frac{\ve}{\eta}\right)R_\eta^H.
\end{equation}
The construction \eqref{general} satisfies Eq. \eqref{lnrho}, provided that
\begin{equation}\label{k}
    k = \frac{as(\ve-\eta)}{\ve \sqrt{1-\eta^2}} =
   \frac{b(\ve-\eta)}{s \eta \sqrt{1-\ve^2}} > 0
\end{equation}
From the $\overline{\rho(x)} = \rho_0$ condition, we find
\begin{equation}\label{rho0}
    \rho_0 = r_0 + \alpha=\frac{s(1-\frac{\eta}{\ve})}{\sqrt{1-\eta^2}}
    + \frac{s \eta \sqrt{1-\ve^2}}{\ve \sqrt{1-\eta^2}}.
\end{equation}

Solving Eq. \eqref{k} and \eqref{rho0}, we find parameters $s$ and $\eta$ as functions of $\ve$ and $\displaystyle{B=\frac{b}{a \rho_0^2}}$. The solutions are
\begin{gather}
    s(B,\ve) = \sqrt{\frac{b\ve}{a\eta}\sqrt{\frac{1-\eta^2}{1-\ve^2}}}
    = \rho_0[1+\frac{\ve^2}{2}B(1-B)+\mathcal{O}(\ve^4)]\,, \label{s} \\
    \eta(B,\ve)=B \ve + \frac{1}{2}B(1-B)^2\ve^3+\mathcal{O}(\ve^5).
\end{gather}
where $\eta$ satisfies the equation
\begin{equation}\label{B}
    B\left[\ve-\eta(1-\sqrt{1-\ve^2})\right]^2=\ve\eta\sqrt{1-\eta^2}\sqrt{1-\ve^2}.
\end{equation}
Parameter $k$ is fixed due to \eqref{k}: 
\begin{equation}
k=a\rho_0-\frac{b}{s}= a \rho_0 \sqrt{B}
    \frac{\ve-\eta}{\sqrt{\ve\eta\sqrt{(1-\ve^2)(1-\eta^2)}}}=
    k_0[1+\frac{1}{2}B^2\ve^2+\mathcal{O}(\ve^4)]. \label{kexp2}
\end{equation}

Let us now discuss some special values of parameters and their implications. For $b=0$ and $a>0$, we have $B=0,\ \eta=0$, and $k = a \rho_0>0$. The solution \eqref{general} reduces to
\begin{equation}
\rho(x) = \rho_0 \frac{\sqrt{1-\ve^2}}{1- \ve \cos kx},
\end{equation}
found in Sec. \ref{sec:3}. For $a=0$, we have $\ve=0,\ \displaystyle{k=-\frac{b}{s}>0}$, and $s=\rho_0\sqrt{1-\eta^2}$. Note that $k>-\frac{b}{\rho_0}$. The corresponding solution reads
\begin{equation}
    \rho(x)= \rho_0 (1- \eta \cos kx).
\end{equation}
Finally, for $a \ne 0$ and $b \ne 0$, we find
\begin{equation}\label{general2}
    \rho(x) =\rho_0+\frac{k}{a}\left(\frac{\sqrt{1-\ve^2}}{1-\ve\cos kx}-1\right).
\end{equation}
In other words, the general solution \eqref{general2} is proportional to a solution \eqref{Reta} shifted by a constant $\displaystyle{\alpha=\rho_o-\frac k a}$.

A couple of limiting cases of solution \eqref{general2} are worth mentioning. If we let $\ve$ tend to zero, we obtain the small $\ve$ expansion \eqref{expansion}, which was introduced in Sec. \ref{sec:3}. Note that, by choosing $C_{1,0}=\dfrac{k_0}{a \rho_0}$, both expansions of $k$ \eqref{kexp} and \eqref{kexp2} are equal up to the second order in $\ve$. 

If we let $k$ tend to zero (or, equivalently, let the period $\displaystyle{\frac{2\pi}{k}} \to \infty$) with $\eta \ne \ve$, and $\eta,\ \ve \to 1$, we obtain aperiodic solutions
\begin{equation}
    \rho(x)= \rho_0 \frac{\dfrac{2(1-\eta)}{k^2}+x^2}{\dfrac{2(1-\ve)}{k^2}+x^2}.
\end{equation}
Let us note that, if $k \le 0,\ \rho(x)=\rho_0.$

We will clarify and analyze these solutions in more detail in the next section.


\section{\label{sec:5}Solutions of two coupled BPS equations}

Having found the solution of Eq. \eqref{BPS1a}, we are now in a position to find the pair of solutions $\rho_1(x)$ and $\rho_2(x)$ of coupled BPS equations \eqref{BPS1} and \eqref{BPS2}. It is important to take into account the normalization conditions
\begin{align}
\overline{\rho_1(x)} &= \rho_{10}=\lim_{L \to \infty} \frac{N_1}{2L} \,, \\
\overline{\rho_2(x)} &= \rho_{20}=\lim_{L \to \infty} \frac{N_2}{2L}. 
\end{align}
which follow from the constraints \eqref{normalize}. This means that both averages $\rho_{10}$ and $\rho_{20}$ are fixed. Namely, it is important to note that the numbers of particles $N_1$ and $N_2$ and the length of the system $L$ are simultaneously taken to infinity, keeping the particle densities $\rho_{10}$ and $\rho_{20}$ fixed. Thus from
\begin{align}
    \rho_1(x) &=s_1 \frac{R_\ve(x)}{R_\eta(x)} \,, \\
    \rho_2(x) &=s_2 \frac{R_\eta(x)}{R_\ve(x)}.
\end{align}
we easily get
\begin{equation}\label{c}
c=\rho_1(x)\rho_2(x)=s_1s_2.
\end{equation}
We use the relation \eqref{rho0} to write the following expressions for $\rho_{10}$ and $\rho_{20}$:
\begin{gather}
    \rho_{10}=\frac{s_1}{\sqrt{1-\eta^2}}\bigg[1-\frac{\eta}{\ve}(1-\sqrt{1-\ve^2})\bigg] \label{rho10}\\
    \rho_{20}=\frac{s_2}{\sqrt{1-\ve^2}}\bigg[1-\frac{\ve}{\eta}(1-\sqrt{1-\eta^2})\bigg].\label{rho20}
\end{gather}
Hence, from \eqref{c}, \eqref{rho10} and \eqref{rho20} we obtain
\begin{equation}
c=\rho_{10}\rho_{20}\frac{\sqrt{(1-\ve^2)(1-\eta^2)}}
    {\Big[1-\dfrac{\eta}{\ve}(1-\sqrt{1-\ve^2})\Big]\Big[1-\dfrac{\ve}{\eta}(1-\sqrt{1-\eta^2})\Big]}.
\end{equation}
Therefore, after fixing both $\rho_{10}$ and $\rho_{20}$, the parameters $c$ and $b = \dfrac{2c\pi}{\lambda_1-1}$ become dependent on $\ve$. Hence, Eq. \eqref{B} becomes
\begin{equation}\label{mu}
    \mu\Big[\ve-\eta(1-\sqrt{1-\ve^2})\Big]=\eta-\ve(1-\sqrt{1-\eta^2})
\end{equation}
where
\begin{equation}
\mu=\frac{\rho_{20}}{\lambda_1\rho_{10}}.
\end{equation} 
Note that
\begin{equation}\label{k2}
\begin{split}
k & = a\rho_{10}-\frac{b}{s_1} =\frac{2\pi}{\lambda_1-1}(\lambda_1\rho_{10}-s_2)= \\
& = \frac{2\pi}{\lambda_2-1}(\lambda_2\rho_{20}-s_1)>0,
\end{split}
\end{equation}
implies
\begin{equation}
\frac{s_2}{\lambda_1\rho_{10}}\le1, \quad
\frac{s_2}{\rho_{20}}\le \frac{1}{\mu}.
\end{equation}

Equation \eqref{mu} has two solutions for $\eta$:
\begin{equation}\label{eta}
\frac{\eta_\pm}{\ve}=
\frac{(\mu+1)^2-\mu(\mu+1)\sqrt{1-\ve^2}\pm \sqrt{2\mu(\mu+1)[1-\ve^2-\sqrt{1-\ve^2}]+1}}
{(\mu+1)^2+\mu^2(1-\ve^2)+\ve^2-2\mu(\mu+1)\sqrt{1-\ve^2}},
\end{equation}
with restriction $\eta_\pm<\ve.$

Let us now classify solutions $\rho_1(x)$ and $\rho_2(x)$ with respect to values of parameter $\mu$.

\begin{itemize}
    \item[i)] \textit{Case} $\mu=0,\quad 0 \le\ve\le 1$. From \eqref{k2} and \eqref{eta} it follows that 
\begin{gather}
    \eta_-=0,\ s_1=\rho_{10},\ s_2=\rho_{20}\sqrt{1-\ve^2}=0, \\
    c=\rho_{10}\rho_{20}\sqrt{1-\ve^2}=0 \quad \textrm{and} \quad 
    k=\frac{2\pi\lambda_1\rho_{10}}{\lambda_1-1}>0.
\end{gather}
It is obvious that this case corresponds to $\rho_{20}/\rho_{10} \to 0$, or the $\lambda_1 \to \infty$ limit. In the latter case (the strong coupling limit),
$a= 2\pi,\ b=\dfrac{2\pi c}{\lambda_1-1} \to 0,\ \ve\to1, \ k=2\pi\rho_{10}$, and solutions $\rho_1(x)$ and $\rho_2(x)$ are given by
\begin{gather}
    \rho_1(x)=\rho_{10}\frac{\sqrt{1-\ve^2}}{1-\ve\cos(2\pi\rho_{10}x)}
\to\sum_{n\in\mathbf{Z}}\delta(x-\frac{n}{\rho_{10}})\\
    \rho_2(x)=\rho_{20}[1-\cos(2\pi\rho_{10}x)].
\end{gather}

\item[ii)] \textit{Case} $0<\mu<1,\quad 0 \le\ve\le 1$. Solutions $\rho_1(x)$ and $\rho_2(x)$ are close to constants $\rho_{10}$ and $\rho_{20}$ for $\ve \ll 1$. For $\ve=1$ they are given by
\begin{gather}
    \rho_{1}(x)=\frac{\lambda_1-1}{\lambda_1}\sum_{n\in\mathbf{Z}}\delta
(x-\frac{2\pi}{k_0}n) \label{rho1} \\
     \rho_2(x)=\rho_{20}\frac{(\mu+1)\Big[(\mu+1)^2-1\Big](1-\cos k_0x)}
{\mu\Big[(\mu+1)^2+1\Big]\Big[1-\dfrac{(\mu+1)^2-1}{(\mu+1)^2+1}\cos k_0x\Big]}, \label{rho2}
\end{gather}
where 
\begin{equation}
k_0=\frac{2\pi \lambda_1}{\lambda_1-1}\rho_{10}.
\end{equation}

\item[iii)] \textit{Case} $\mu=1$. The allowed range for $\ve$ is $\sqrt{3}/2 \le\ve\le1$. The function
$\frac{\eta_-}{\ve}$ in that range reads
\begin{equation}
\frac{\eta_-}{\ve}=\frac{3}{5-4\sqrt{1-\ve^2}}.
\end{equation}

In the interval $0\le\ve\le\sqrt{3}/2$, $\eta_-=\ve$, while for the interval
$\sqrt{3}/2\le\ve\le1,\ \eta_+=\ve$. Hence solutions
$\rho_1(x)$ and $\rho_2(x)$ reduce to constants $\rho_{10}$ and $\rho_{20}$, respectively.

For $\ve=1$, the solutions are given by
\begin{gather}
    \rho_{1}(x)=\frac{\lambda_1-1}{\lambda_1}\sum_{n\in\mathbf{Z}}\delta
(x-\frac{2\pi}{k_0}n) \\
     \rho_2(x)=1.2\rho_{20}\frac{(1-\cos k_0x)}{(1-0.6\cos k_0x)}.
\end{gather}

\item[iv)] \textit{Case} $\mu>1$. There are two solutions for $\dfrac{\eta_\pm}{\ve}$ if $\ve_{cr.}\le\ve\le1$,
where 
\begin{equation}
\ve_{cr.}=\frac{1}{2}\sqrt{3+\sqrt{1-\frac{2}{\mu(\mu+1)}}}.
\end{equation}
At $\ve=\ve_{cr.}$ we find $\eta_+=\eta_-<\ve_{cr.}$.
At $\ve=1$ we have
\begin{equation}
\frac{\eta_-}{\ve}=\frac{(\mu+1)^2-1}{(\mu+1)^2+1} \quad \textrm{and} \quad 
\frac{\eta_+}{\ve}=1.
\end{equation}

\end{itemize}

For $\eta_-=\frac{(\mu+1)^2-1}{(\mu+1)^2+1},\ s_1=\rho_{10}(\mu+1),\ s_2=0,\ c=0$, and $k_0=\frac{2\pi\lambda_1}{\lambda_1-1}\rho_{10}$, solutions $\rho_1(x)$ and 
$\rho_2(x)$ are periodic of type \eqref{rho1}, \eqref{rho2}.

The solutions $\rho_1(x)$ and $\rho_2(x)$ are aperiodic for $\eta_+\to1,\ \ve\to1,\
s_1=\mu\rho_{10},\ s_2=\frac{\rho_{20}}{\mu},\
c=\rho_{10}\rho_{20},\ \mu=\sqrt{(1-\eta^2)/(1-\ve^2)}$, and $k\to0_+$ and reduce to
\begin{gather}
    \rho_1(x)=\rho_{10}\frac{\dfrac{2(1-\eta)}{k^2}+x^2}{\dfrac{2(1-\ve)}{k^2}+x^2} \label{aperiodic1}\\
    \rho_2(x)=\rho_{20}\frac{\dfrac{2(1-\ve)}{k^2}+x^2}{\dfrac{2(1-\eta)}{k^2}+x^2} \label{aperiodic2}
\end{gather}

These solutions were already described in Refs. \cite{BMEPL:2005} and \cite{BM:2006}. For $\lambda_1<1$, the first solution \eqref{aperiodic1} behaves like the hole in the condensate $\rho_{10}$, and the second one \eqref{aperiodic2} behaves like the particle above the condensate $\rho_{20}$. The roles are interchanged for $\lambda_1>1$.

We point out that there are no other aperiodic solutions which can appear from
our construction $\rho_1(x)=s_1 \dfrac{R_\ve(x)}{R_\eta(x)}$ and
$\rho_{2}(x)=\dfrac{c}{\rho_{1}(x)}$.


\section{\label{sec:6}Discussion and conclusion}

The perturbative method developed in Sec. \ref{sec:3} can be applied to a wide class of coupled BPS equations. In the general case for $\lambda_1 \lambda_2=\lambda_{12}^2 \ne 1$, it can be shown that the coupled BPS equations \eqref{BPS1} and \eqref{BPS2} lead to the following relation between $\rho_1$ and $\rho_2$:
\begin{equation}
(\lambda_1-1)\lambda_{12}\frac{\rho_1'}{\rho_1}=(\lambda_2-1)\lambda_{1}\frac{\rho_2'}{\rho_2}.
\end{equation}
This means that $\rho_2(x) = \tilde c \rho_1^\kappa(x)$,
where 
\begin{equation}
\kappa={\frac{(\lambda_1-1)\lambda_{12}}{(\lambda_2-1)\lambda_{1}}}.
\end{equation}
Hence, the most general coupled BPS equations reduce to
\begin{equation}
(\ln \rho_1)'=a\rho_1^H + b(\rho_1^\kappa)^H,\quad \rho_1(x)\ge0,\
\overline{\rho_1(x)}=\rho_{10}
\end{equation}
where $a = \dfrac{2\pi\lambda_1}{\lambda_1-1}$ and $b = \dfrac{2\pi\lambda_{12}}{\lambda_1-1}\tilde c $. The solutions of this equation possess the following general properties:
\begin{itemize}
    \item[i)] There are constant solutions $\rho_1(x)=\rho_{10}=$ const.

    \item[ii)] Solutions close to constant $\rho_{10}$:
\begin{equation}
\rho_1(x)=\rho_{10}+\ve\vp_1(x),\quad |\ve\vp_1(x)|\ll\rho_{10},\ \overline{\vp_1(x)}=0
\end{equation}
where $\vp_1(x)$ satisfies
\begin{equation}
\vp_1'=(a\rho_{10}+\kappa b\rho_{10}^\kappa)\vp_1^H
\end{equation}
and $k_0=a\rho_{10}+\kappa b\rho_{10}^\kappa >0$. It is easy to see that the solution $\vp_1$ is given up to a phase by
\begin{equation}
\vp_1(x)=c_1\cos k_0x.
\end{equation}

     \item[iii)] Generally, there exist one-parameter $\ve$ solutions containing constant solution $\rho_{10}$. 

    \item[iv)] There are periodic solutions if $k>0$.

    \item[v)] Aperiodic solutions appear if $k\to0_+$.
\end{itemize}

Our perturbative method allows one to calculate higher order terms in the expansion \eqref{expansion}. However, we have not found the solutions in closed analytic form for $\lambda_{12}^2 \ne 1$. We have found the exact solutions only in two cases: $\lambda_{12}=+1$ and $\lambda_{12}=-1$.

In the case $\lambda_1\lambda_2=1,\ \lambda_{12}=1$, we have $\kappa=-1$,
and in Sec. \ref{sec:5} we have found and classified all solutions of the form:
\begin{gather}
    \rho_1(x)=(\rho_{10}-\lambda_2s_2)R_\ve+\lambda_2s_2\ge0\\
    \rho_2(x)=(\rho_{20}-\lambda_1s_1)R_\eta+\lambda_1s_1\ge0,
\end{gather}
where $\rho_1(x)\rho_2(x)=s_1s_2=c\ge0$ and $s_1,\ s_2$, and $\eta$ are given by Eqs. \eqref{rho10}, \eqref{rho20}, \eqref{eta}, respectively.

Finally, we point out that, in the special case $\lambda_1\lambda_2=1,\ \lambda_{12}=-1$, and $\kappa=1$, the new exact duality appears
with nice properties and a physical interpretation connecting particles and antiparticles  \cite{BFM:2007}, \cite{Sergeev}, \cite{BFM:2008}.

In conclusion, we have studied the two-family Calogero model on line in the
limit in which each family contains a large number of particles. We have 
found that, in the strong-weak dual case, there exists only one nonperiodic soliton-antisoliton, topological
solution \eqref{aperiodic1} and \eqref{aperiodic2}, and periodic, stationary waves solutions \eqref{general2}. Our collective-field approach can be analogously applied to the two-family Sutherland model on a circle of perimeter length $L$.
However, the Hilbert transform must be modified in order to take into
account the compact support of the Sutherland model. Namely, the 
standard kernel $P\frac{1}{x-y}$ should be replaced by the $\cot \frac{\pi}{L}(x-y)$ kernel \cite{Abanov}.
We hope to report on these issues in a separate publication.

\begin{acknowledgments}
This work was supported by the Ministry of Science and Technology of the Republic of Croatia under Contract No. 098-0000000-2865.
\end{acknowledgments}


\end{document}